\documentclass[12pt,preprint]{aastex}
\usepackage{natbib}
\bibpunct[]{(}{)}{,}{a}{}{,}

\providecommand{\km}{{\rm km}}
\providecommand{\cm}{{\rm cm}}
\providecommand{\g}{{\rm g}}
\providecommand{\dyn}{{\rm dyn}}

\providecommand{\s}{{\rm s}}

\shorttitle{KBO Size Spectrum} 

\begin{document}

\title{Shaping the Kuiper Belt Size Spectrum
by Shattering Large but Strengthless Bodies}

\author{Margaret Pan and Re'em Sari}

\affil{130-33 Caltech, Pasadena, CA 91125}

\begin{abstract}
The observed size distribution of Kuiper belt objects (KBOs)---small
icy and rocky solar system bodies orbiting beyond Neptune---is well
described by a power law at large KBO sizes. However, recent work by
\cite{bernstein03} indicates that the size spectrum breaks and becomes
shallower for KBOs smaller than about 70~km in size. Here we show that
we expect such a break at KBO radius $\sim$40~km since destructive
collisions are frequent for smaller KBOs. Specifically, we assume that
KBOs are rubble piles with low material strength rather than solid
monoliths. This gives a power-law slope $q\simeq 3$ where the number
$N(r)$ of KBOs larger than a size $r$ is given by $N(r) \propto
r^{1-q}$; the break location follows from this slope through a
self-consistent calculation. The existence of this break, the break's
location, and the power-law slope we expect below the break are
consistent with the findings of \cite{bernstein03}. The agreement with
observations indicates that KBOs are effectively strengthless rubble
piles.
\end{abstract}

\section{Introduction}

The Kuiper belt, a population of small bodies moving beyond the giant
planets, was discovered when its first member was found in 1992
\citep{jewitt93}.  As of late 2003, $\sim$800 KBOs have been
discovered. Due to KBOs' faintness, however, the size distribution of
KBOs is well determined observationally only for bodies larger than
$\sim$100~km \citep{trujillo01,gladman98,chiang99}; their size
spectrum is consistent with a power law $N(r)\propto r^{-4}$
\citep{bernstein03}. Numerical studies concluded that the differential
size spectrum below $\sim$100~km should follow a power law with the
slightly shallower $N\propto r^{-2.5}$ due to the effects of
destructive collisions \citep{farinella96,davis97,kenyon02}.  The
results seemed consistent with loose observational constraints
available on the number of $\sim$20~km and $\sim$2~km KBOs
\citep{cochran95,holman93}.

In this context, the deficit in small KBOs observed by
\cite{bernstein03} was a surprise. Using the Advanced Camera for
Surveys recently installed on the Hubble Space Telescope, they found
just 3 KBOs of size $\sim$25--45~km where they expected $\sim$85 such
bodies based on an extrapolation of the accepted best-fit large-KBO
spectrum at the time \citep{trujillo01}. While this observed decrement
of more than an order of magnitude in the number of small KBOs clearly
indicates a break between 45 and 100~km, the exact break position and
slope below the break may well be refined by future data on small
KBOs. Still, the results of \cite{bernstein03} are inconsistent with
the previously expected small-end spectrum $N\propto R^{-2.5}$, or
$q=3.5$, at better than 95\% confidence.

This paper describes a simple self-consistent analytic calculation of
the break location and the slope below the break. Note that using the
$N(r)\propto r^{-4}$ size spectrum obtained by \cite{bernstein03} for
large KBOs, we can estimate the size below which collisions between
equal size bodies should be frequent to be $\sim$ 1~km---well below
the observed break location. However, this estimate needs two
modifications.  First, due to the large velocity dispersion in the
Kuiper belt, small bodies can shatter much larger objects. Since there
are more small than large bodies, destructive collisions will occur
frequently even for objects much larger than 1~km. Second, when
collisions are important, they reduce the number of small bodies; this
in turn decreases the frequency of collisions. Therefore, calculations
of the effects of collisions and the size below which collisions are
important must be done in a self-consistent manner.

\section{Slope of the steady-state distribution}

In order to find the break location self-consistently, we first
calculate the power-law slope $q$ for a collisional population of
bodies. We assume a group of bodies with isotropic velocity dispersion
$v$ in which the differential number of bodies of radius $r$ is given
by a power law $dN(r)/dr \propto r^{-q}$. If we assume that the
population is in a steady state and that mass is conserved in the
collision process, the total mass of bodies destroyed per unit time in
a logarithmic interval in radius must be independent of size. We can
use this condition to determine $q$.

\begin{figure}
\epsscale{.7}
\plotone{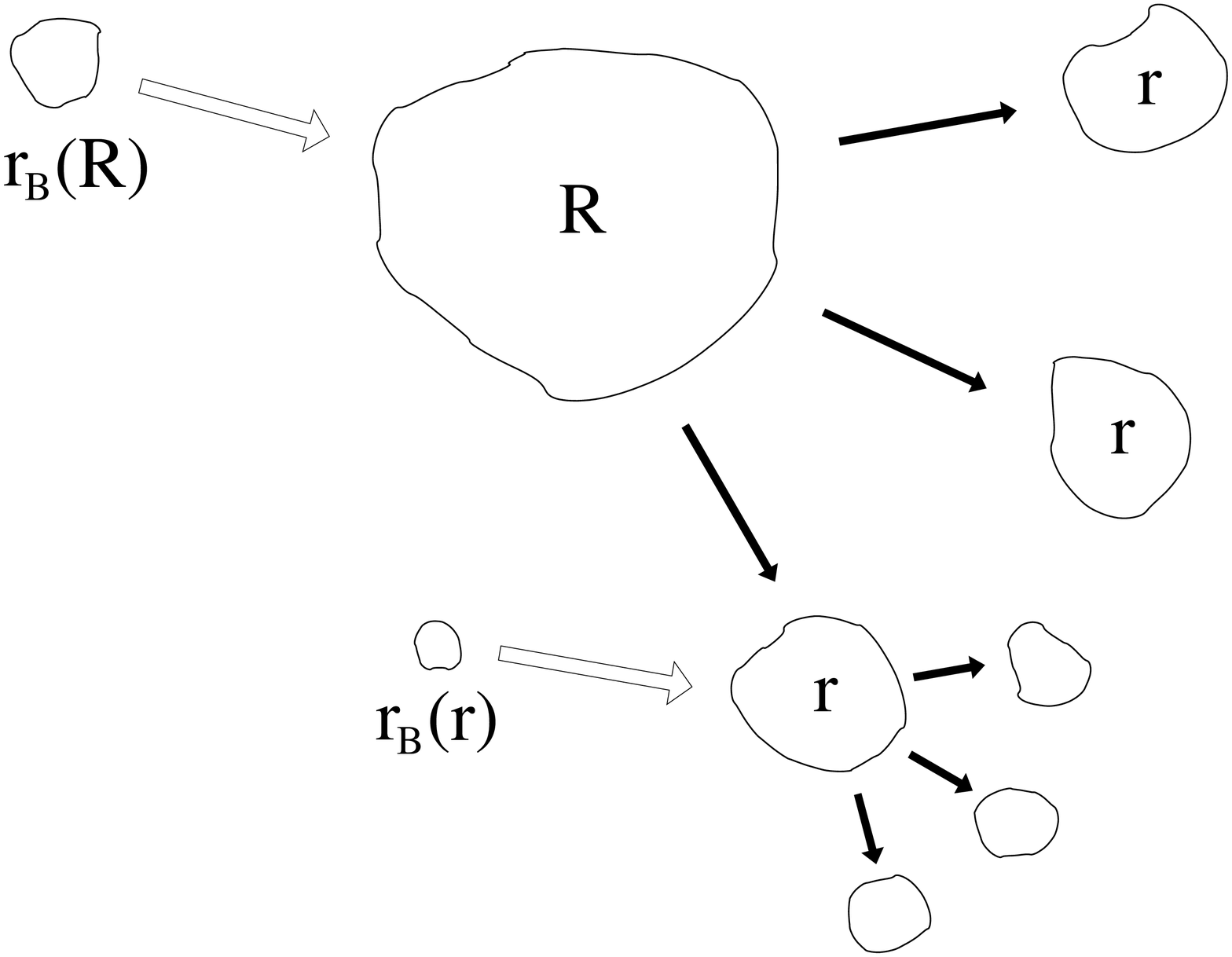}
\caption{Schematic of the collisional cascade: bullets of size
$r_B(R)$ shatter targets of typical size $R$; these targets break into
new targets of size $r$, which are in turn shattered by bullets of
size $r_B(r)$; and so on. Since mass is conserved in collisions, the
mass destruction rate of bodies of size $R$ is the mass creation rate
of bodies of size $r$. Steady state then requires that the rate of
mass destruction be independent of body size.}
\label{cascade}
\end{figure}

We assume that the main channel for mass destruction is the shattering
of larger `targets' by smaller `bullets' (Fig.~\ref{cascade}). Under
this condition,
\begin{equation}
\label{steadystate}
\rho r^3\cdot N(r)\cdot \frac{N(r_B)}{V} \cdot r^2\cdot v = \rm{constant}
\;\;\; .
\end{equation}
Here $\rho$ is the internal density of each body and $r_B(r)$ is the
size of the smallest bullet which, on impact, can shatter a target of
radius $r$. $V$ is the volume occupied by all the bodies; their
velocity dispersion and therefore their distribution within $V$ are
assumed independent of body size.  When supplemented by a relation
between the size of the bullet and that of the target,
Eq.~\ref{steadystate} dictates the size spectrum $q$.

This very simple formalism based on conservation of mass captures the
essence of Dohnanyi's (\citeyear{dohnanyi69}) more elaborate
pioneering treatment. Based on laboratory experiments which involved
solid bodies dominated by material strength, Dohnanyi chose
$r_B\propto r$. When $r_B\propto r$ and $N(r)\propto r^{1-q}$ are
inserted into Eq.~\ref{steadystate}, we retrieve the $q=7/2$ of
Dohnanyi and several subsequent authors \citep[for example,
][]{williams94,tanaka96}. This slope is much steeper than the best-fit
small-end $q=2.3$ found by \cite{bernstein03}, who rule out $q=7/2$ at
better than 95\% confidence.

Indeed, work on the structure of small solar system bodies suggests
that many of them are gravitationally bound rubble piles rather than
solid monoliths. Based on oscillating lightcurves of the large KBO
(20000) Varuna (radius $R>100\;\km$), \cite{jewitt02} find that this
bodies has density $\sim 1\;\g\;\cm^{-3}$ and is therefore unlikely to
be solid. However, other effects may also be responsible for the
lightcurve shape \citep[see, for example,][]{goldreich04}. The
rotation statistics of much smaller bodies ($R\sim 10\;\km$) in the
more easily observed region between the asteroid belt and the sun also
suggest that small bodies in the solar system are rubble piles rather
than monoliths.  That no small asteroids are observed to rotate faster
than their breakup speed suggests that those which were spun up beyond
breakup simply broke apart \citep{harris96}; this in turn suggests
that these asteroids have no tensile strength. A study including 26
small near-earth asteroids came to similar conclusions about asteroid
internal structure \citep{pravec98}. The most detailed probe available
of the structure of small KBOs is research on short-period comets,
kilometer-sized bodies which are thought to have originated in the
Kuiper belt. Work on the breakup and impact of comet Shoemaker-Levy 9,
thought to be 1--2~km in size, indicates that its strength before
breakup was $\sim60\;\dyn\;\cm^{-2}$ or less \citep{asphaug96}; a body
like Shoemaker-Levy 9 would have material strength energy at most
about ten times less than its gravitational energy. These indications
motivate an investigation of the influence of negligible material
strength on the fragmentation spectrum.

We might therefore replace the $r_B\propto r$ destruction criterion
used by Dohnanyi with the requirement that the kinetic energy of the
bullet be equal to the total gravitational energy of the target:
\begin{equation}
\label{equalenergies}
\rho r_B^3 v^2 \sim \rho r^3 v_{\rm esc}^2
\end{equation}
where $v_{\rm esc}\propto r$ is the escape velocity from a target
of size $r$, and, again, $v$ is the bodies' constant velocity
dispersion. Then
\begin{equation}
\label{destruct_cond}
r_B(r) \sim \left(\frac{G\rho}{v^2}\right)^{1/3} r^{5/3} 
       \sim r_{\rm eq}^{-2/3} r^{5/3} \;\;\; , \;\;\;
r_{\rm eq} \sim \frac{v}{\sqrt{G\rho}} \;\;\; .
\end{equation}
Physically, $r_{\rm eq}$ is the size of a body whose escape velocity
equals the velocity dispersion of the system. When density $\rho\sim
1\;\g\;\cm^{-3}$ and the Kuiper Belt's current velocity dispersion
$v\sim 1\;\km\;\s^{-1}$ are used, $r_{\rm eq} \sim 10^3\;\km \sim$ the
radius of Pluto. Equivalently, a target of size $r_{eq}$, or roughly
Pluto's size, would require a bullet of equal mass to shatter it. Then
a body smaller than Pluto---that is, virtually any KBO---can be
shattered by bullets smaller than itself.  When we substitute
Eq.~\ref{destruct_cond}, or essentially the proportionality $r_B
\propto r^{5/3}$, into Eq.~\ref{steadystate}, we get the power-law
slope
\begin{equation}
\label{q}
q=23/8 \;\;\; . 
\end{equation}
Recently, \cite{obrien03} extended Dohnanyi's treatment to other
destruction conditions where $r_B$ scales as an arbitrary power of
$r$; they show that $q$ is a simple function of this power so that a
range of $q$ values can be obtained from a calculation like
Dohnanyi's. The simple argument we express in Eq.~\ref{steadystate}
reproduces their analytic results for $q$. Eq.~\ref{destruct_cond} is
a special cases of their general power law which is clearly motivated
by energy considerations and which leads to the spectrum given by
Eq.~\ref{q}.

\section{Realistic destruction criteria}

The destruction criterion just discussed neglects any energy loss
during the impact process. It is then a lower limit on the energy
needed to shatter and disperse a given target. Indeed, numerical
simulations and dimensional analysis of impact events find that in the
`gravity regime', or target size range where gravity dominates
material strength, the impact energy needed to shatter a given target
lies well above the level indicated by Eq.~\ref{destruct_cond}
\citep{housen90,holsapple94,love96,melosh97,benz99}. Further, the
$r_B(r)$ scalings indicated by these studies%
\footnote{Again, we assume a constant velocity dispersion for the
collisional population.}  
are consistently shallower than the one in
Eq.~\ref{destruct_cond}. With $r_B\propto r^\alpha$, they give
$1.37\leq \alpha \leq 1.57$ rather than the $\alpha=5/3$ in
Eq.~\ref{destruct_cond}.

Upon insertion into Eq.~\ref{steadystate}, the $r_B(r)$ scalings above
give $2.95<q<3.11$. These values indicate a spectrum
between the one given by Eq.~\ref{q} and Dohnanyi's $q=3.5$. %
This range in $q$ is consistent with the best-fit slope $q=2.8\pm 0.6$
(95\% confidence) derived by \cite{bernstein03} below the break for
the classical Kuiper belt and with the best-fit $q=2.3^{+0.9}_{-1.1}$
(bounds of 68\% confidence contour) slope they find for the entire
Kuiper belt.  The value for the entire belt may be skewed downwards by
the scattered Kuiper belt data, which include too few faint objects
for the scattered belt's small-end slope to be well determined.  The
observed KBO spectrum is thus consistent with the assumption that
gravity dominates material strength in KBOs of size near $r_{\rm
break}$.

That the simulations give $r_B(r)$ scalings shallower than that of
Eq.~\ref{destruct_cond} implies that the energy lost in a catastrophic
collision depends on the bullet/target size ratio. As has previously
been noted \citep[see, for example,][]{melosh97}, we would expect
energy loss in the impact of a small bullet on a much larger
target. Initially the bullet would transfer most of its energy to a
volume the size of itself at the impact site; much of this energy
would escape from the site via a small amount of fast ejecta, though
some would propagate through the target as a shock.

Somewhat more quantitatively, we can think of a collision between a
very small bullet and a large target as a point explosion on the
planar surface between a vacuum and a half-infinite space filled with
matter. The analogous explosion in a uniform infinite material leads
to the Sedov-Taylor blast wave, a self-similar solution of the first
type in which total energy is conserved as the spherical shock
propagates \citep{sedov46,taylor50}. By contrast, a point explosion in
a half-infinite space is a self-similar solution of the second type
\citep{zeldovich67}; the shock moving into the half-space must lose
energy as some of the shocked material flows into the vacuum. Also,
the nonzero pressure increases the momentum in the shock. So as the
shock propagates, its velocity should fall off faster than it
would have given conservation of energy but slower than it would have
in the case of momentum conservation.

We can use these considerations to constrain $r_B(r)$ scalings for
catastrophic collisions. We assume that a given target is destroyed if
the velocity of the shock wave when it reaches the antipode of the
impact site exceeds the escape velocity \cite[see, for
example,][]{melosh94}. Let the shock velocity decay as $v_{\rm
shock}\propto x^{-\beta}$ where $x$ is the distance traveled by the
shock. If the energy in the shock were conserved, we would expect
$\beta=3/2$ from dimensional analysis; if the momentum were conserved,
we would expect $\beta=3$. Then the actual point explosion solution
must have $3/2<\beta<3$. The criterion for target destruction is
\begin{equation}
\label{destruct_cond_shock}
\rho r_B^3 v^2 \left(\frac{r}{r_B}\right )^{3-2\beta} \sim G\rho^2 r^5
\end{equation}
where we have assumed the bullet initially deposits its energy in a
volume the size of itself. This implies
\begin{equation}
r_B\propto r^{1+1/\beta} \;\;\; , \;\;\; q=\frac{7\beta +1}{2\beta +1}
\;\;\; .
\end{equation}
The $3/2<\beta<3$ condition requires $4/3 < \alpha < 5/3$ and
$23/8<q<22/7$, both of which are satisfied by all of the impact
simulation and dimensional analysis results. \citet{holsapple94}
mentions that energy and momentum conservation should represent
limiting cases for the impact process and that laboratory experiments
involving impacts into sand, rock, and water satisfy those
limits. Note that the range in $q$ found in previous studies,
$2.95\leq q\leq 3.11$, spans most of the allowed range for $q$. This
suggests that the catastrophic impact process and $\alpha$ depend on
more specific details of the collisions such as the equation of state.

At $r_B\sim r$ there should be no energy loss because the initial
energy is deposited in a volume of linear size
$r$. Eq.~\ref{destruct_cond_shock} reflects this. Then the $r_{\rm eq}$ 
expression in Eq.~\ref{destruct_cond} is still valid.

\section{Location of the break}

The above calculation of the size spectrum treats $N(r)$ as constant
in time. To maintain this steady-state exactly would require the power
law to extend to bodies of infinite size, which is impossible.  To
find the range of masses where this assumption holds, we first find
the size $r_{\rm break}$ of the largest KBO to have experienced a
destructive collision after an elapsed time $\tau$.  We equate the
timescale for destructive collisions for each KBO of size $r_{\rm
break}$ to $\tau$ using Eqs.~\ref{steadystate} and
\ref{destruct_cond_shock}. To get $N(r)$ we note that bodies of size
$r>r_{\rm break}$, having never collided, should be effectively
primordial at time $\tau$. For their size spectrum we write $N(r)=N_0
r^{1-q_0}$ where $N_0\sim 4\times 10^{7q_0-3}\;\cm ^{q_0-1}$ from
observations \citep{trujillo01}. This is equivalent to a Kuiper belt
with $4\times 10^4$ bodies larger than 100~km. They are spread over an
area $A\simeq 1200\;{\rm AU}^2$ in the plane of the solar system
\citep{trujillo01}, so $V\simeq Av/\Omega$ where $\Omega=0.022\;{\rm
yr}^{-1}$ is the typical orbital angular velocity of the Kuiper
belt. With $q$ for the slope below the break and, as above, $q_0$ and
$N_0$ for the slope and normalization above the break, we have
\begin{equation}
\label{rbreakexp}
r_{\rm break} 
\sim \left[\frac{N_0 \Omega \tau}{A} r_{\rm eq}^{7-2q}\right] 
     ^{\frac{1}{4+q_0-2q}}
\end{equation}
If we set $\tau\simeq 4.5\times 10^9\;{\rm yr}$ to be the age of the
solar system, take $3/2<\beta<3$, and use the observed $q_0=5$, we get
\begin{equation}
\label{rbreak}
20\;\km \lesssim r_{\rm break} \lesssim 50\;\km \;\;\; .
\end{equation}
This is consistent with the observed break position of $\sim$70~km
\citep{bernstein03}. Note that if the system had had the high velocity
dispersion assumed above over a time considerably shorter than
4.5~Gyr, the break would have occurred at a much smaller KBO size. We
therefore infer that the Kuiper belt's current excited state has been
a long-lived phase of at least a few billion years' duration rather
than a recent phenomenon.

The evolution of the total mass and velocity dispersion of the Kuiper
belt is a potential concern, as the break location depends strongly on
both.  The mass of the Kuiper belt may have been larger by a factor of
$\sim$100 when the solar system was very young ($10^7-10^8$ years old)
\citep[see, for example,][]{kenyon02}.  The collision frequency would
have been much higher then, so collisions during that period might be
expected to have increased the break radius.  At that time, though,
the velocity dispersion of KBO precursors is believed to have been
just $\sim$1~m/s \citep[see, for example,][]{goldreich02}. With this
impact velocity, $r_{\rm eq} \sim 1\;\km$, so only targets of size
$<$1~km can be shattered by bullets smaller than they. As a result,
collisional evolution during the early solar system should only have
affected bodies of size $<$1~km.  The observed break in the spectrum
must have been created later.  The break location could have been
affected if there was a sufficiently long period during which both $v$
and the Kuiper belt mass were large.

\begin{figure}[h!]
\epsscale{.7}
\plotone{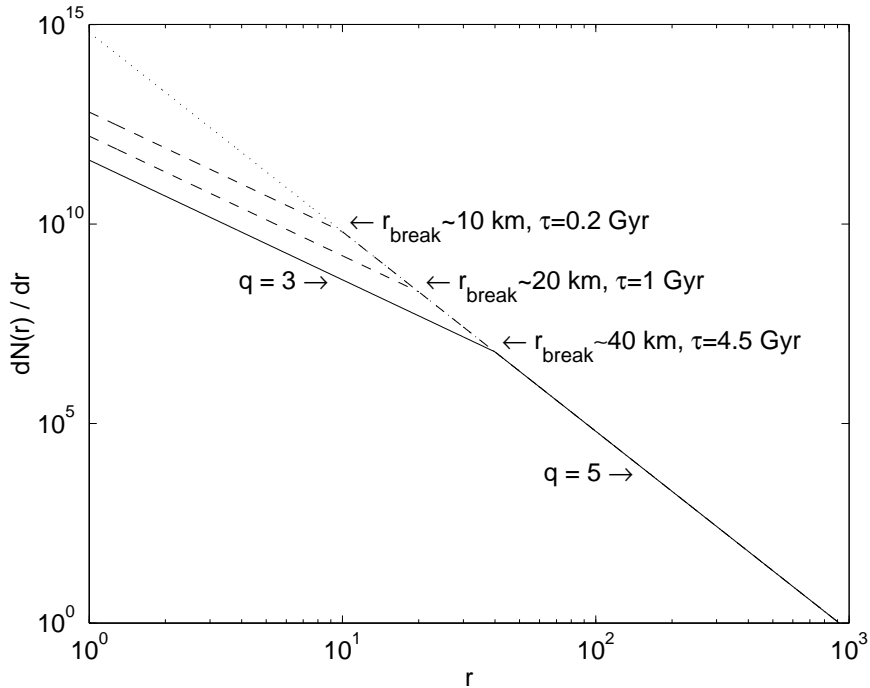}
\caption{Temporal evolution of the number of bodies. Here we use
$q=3$ as a numerical example. The solid line represents the current
KBO size distribution. The dotted line is the extrapolation of the
large-KBO spectrum to small sizes; we assume this line also represents
the primordial size spectrum. Dashed lines show the size spectrum at
earlier times $\tau=0.2$ and 1 billion years. Because $r_{\rm break}$
increases with time, $N(r_{\rm break})$ decreases with time. The
evolution of $r_{\rm break}$ and $N(r_{\rm break})$ is much slower
than the rates of collisional destruction and creation below $r_{\rm
break}$, so these two rates must be very nearly in balance. Then the
steady-state approximation is valid in this size range and the
spectrum below $r_{\rm break}$ follows a $q=3$ power law. }
\label{spec_evol}
\end{figure}

With $q\simeq 3$, the mass contained in bodies of size $r \ll
r_{break}$ is $N(r)\rho r^3\propto r$. Since the mass destroyed per
unit time is independent of body size, the timescale on which
collisional equilibrium is established is $(r/r_{\rm break})\tau \ll
\tau$. Then the steady-state approximation---our assumption that the
rate at which $N$ changes is much less than the rate of destructive
collisions---is self-consistent for $r \ll r_{\rm
break}$. Specifically, as $r_{\rm break}$ increases, $N(r_{\rm
break})$ decreases---both on a timescale $\tau$---and the $q\simeq 3$
spectrum below $r_{\rm break}$ follows adiabatically
(Fig.~\ref{spec_evol}).  Our formalism yields the asymptotic spectrum
far below $r_{\rm break}$ even though the system is not in steady
state overall, since for $r\ll r_{\rm break}$ the destruction rate is
faster than the evolution timescale of the system. \cite{dohnanyi69}
did not discuss the slow decrease in $N$ by which the spectrum differs
from a true steady state; he claimed that non-steady-state power-law
solutions do not exist. \cite{bernstein03} conjectured that the
disagreement between their results and Dohnanyi's calculations might
indicate a non-steady-state condition in the Kuiper belt. However, the
discussion above shows that the fragmentation spectrum below $r_{\rm
break}$ should be unaffected by the system's evolution.

As for the lower size boundary, the strength limit derived by
\cite{asphaug96} implies that material strength dominates gravity
at $r\lesssim0.3\;\km$. Impact simulations reach similar conclusions;
they put the threshhold in the 0.1--1~km size range \citep[see, for
example,][]{love96,melosh97,benz99}. Below this size threshhold a
different $q$ will apply to an equilibrium collisional population.
The changes introduced by this effect in the KBO size distribution
below $\sim$100~m will affect the size distribution of larger bodies
through catastrophic collisions. Numerical simulations of collisional
populations indicate that `waves' may appear in the size spectrum due
to a break in the spectrum introduced by a different $q$
\citep{obrien03}. However, the simulations indicate that the average
slope of the `wavy' spectrum is not affected; also, the distribution
should asymptotically approach a $q\simeq 3$ spectrum far above this
size.

\section{Summary}

We have derived a self-consistent size spectrum $23/8<q<22/7$ for a
collisional population of bodies whose binding energy is dominated by
gravity. We emphasize that this spectrum does not truly represent a
steady state; instead, the number density of bodies decreases slowly
compared to the collision timescale. For the case of the Kuiper Belt,
the spectrum's small-end power-law slope $q\simeq 3$ and break radius
$r_{\rm break}\sim 40\;\km$ agree well with those found
observationally by \cite{bernstein03}. Since the power-law slope
derived in the steady-state approximation depends heavily on the
particular criterion for catastrophic destruction adopted for the
bodies, observations of the KBO size spectrum provide a direct
constraint on the bodies' internal structure.  The close agreement
between this slope and break radius and the best-fit values found by
\cite{bernstein03} suggests that large KBOs are virtually strengthless
bodies held together mainly by gravity.  Further surveys of small KBOs
between $\sim$10 and $\sim$70~km in size would better constrain both
the exact position of the actual break in the size spectrum and the
power-law slope below the break. Data of this kind would thus confirm
or refute our analysis. Such surveys would also allow more detailed
comparison of the break locations in the classical and scattered KBO
populations, which should reflect differences in the surface densities
and velocity dispersions for those two groups.

\noindent {\em Acknowledgements.} We thank Oded Aharonson and Andrew
McFadyen for useful discussions. MP is supported by an NSF graduate
research fellowship.

\bibliographystyle{apj}
\bibliography{kbospec}

\end{document}